\documentclass[paper]{JHEP3} 

\usepackage{graphicx}
\usepackage{latexsym}
\usepackage[abs]{overpic}
\usepackage{wrapfig}
\usepackage{amssymb}
\usepackage{amsmath}

\newcommand{\eq} {equation}
\newcommand{\eqa} {eqnarray}
\newcommand{\NN} {\mbox {$\nonumber$}}
\newcommand{\del} {\mbox {$\partial$}}

\def\Tr{\mathrm{Tr}}

\def\R{{\mathbb{R}}}
\def\Z{{\mathbb{Z}}}

\def\slashb#1{\not\!\!#1}

\newcommand{\be}{\begin{equation}}
\newcommand{\ee}{\end{equation}}
\newcommand{\bea}{\begin{eqnarray}}
\newcommand{\eea}{\end{eqnarray}}

\allowdisplaybreaks
\title{
\huge Maximal super Yang-Mills theories on curved background with off-shell supercharges
}

\author{
Masashi Fujitsuka$^{a}$, Masazumi Honda$^{a,b}$ and Yutaka Yoshida$^b$
\vspace*{0.5cm} \\
\llap{$^a$}Department of Particle and Nuclear Physics,\\
Graduate University for Advanced Studies (SOKENDAI),\\
Tsukuba, Ibaraki 305-0801, Japan\\
\llap{$^b$}High Energy Accelerator Research Organization (KEK),\\
Tsukuba, Ibaraki 305-0801, Japan\\
\vspace*{0.5cm} \\
\email{fmasashi@post.kek.jp, mhonda@post.kek.jp, yyoshida@post.kek.jp}}

\preprint{\footnotesize KEK-TH-1570}

\abstract{We construct $d\leq 7$ dimensional maximally supersymmetric Yang-Mills theories 
on a class of curved backgrounds with off-shell supercharges.
The off-shell supersymmetry is mainly a generalization of on-shell supersymmetry 
previously constructed  by Blau. 
We present several examples of backgrounds and discuss the number of the preserved supersymmetries on these backgrounds.  
We also construct another maximally supersymmetric Yang-Mills theories on $S^3$ 
obtained by dimensional reduction along $\mathbb{R}$-direction of $\mathcal{N}=4$ super Yang-Mills theory on $\mathbb{R} \times S^3$. 
}


\begin{document}

\tableofcontents

\section{Introduction}
The maximally supersymmetric Yang-Mills theories (SYM) have prominent structures at quantum level and
deep connections with superstring theory \cite{Maldacena:1997re,Itzhaki:1998dd}. 
One of most typical examples is the four-dimensional $\mathcal{N}=4$ SYM.
Since the beta function of this theory vanishes, 
the theory is known  as conformal field theory \cite{Avdeev:1980bh,Grisaru:1980nk,Caswell:1980ru,Sohnius:1981sn,Howe:1983sr}. 
Furthermore, it has been also proven that the theory is UV-finite \cite{Howe:1983sr,Mandelstam:1982cb,Brink:1982wv}.
In spite of such simple UV structures, 
the supersymmetric transformations for the component fields seem much more
complicated than less supersymmetric gauge theories.
This is because the supersymmetry closes only on-shell and is not realized linearly.
Actually we do not know whether fully off-shell formalism of the maximal SYM exists or not 
and this has been the one of long-standing problems in theoretical physics.
In ref.~\cite{Brink:1982pd},
the authors have constructed
the light-cone gauge fixed action of the four-dimensional $\mathcal{N}=4$ SYM, 
where eight supersymmetries close off-shell in this formalism.
While this is quite useful for proofs of the UV finiteness \cite{Brink:1982wv,Ananth:2006ac,Ananth:2007px},
the action is non-local and does not have manifest Lorentz and gauge invariance.
In 1993, Berkovits has constructed the local off-shell supersymmetric action 
with manifest Lorentz and gauge invariance by introducing seven auxiliary fields \cite{Berkovits:1993zz}. 
One can show that the maximal number of the off-shell supercharges is nine \cite{Berkovits:1993zz,Evans:1994np}.

Off-shell supersymmetry plays an important role in theoretical physics.
One of most important examples is that
an off-shell  supercharge in a theory
enables us to apply localization method \cite{Pestun:2007rz} to the theory in principle,
which often drastically simplifies its analysis. 
However, even if we construct the off-shell supersymmetry in the flat space and 
apply the localization procedure,
the partition function or correlation functions often suffer from divergences coming from the infrared effect 
and flat directions. 
The former is regularized by putting the theory on compact spaces and the latter is regularized  by introducing mass terms.
Pestun has solved these problems simultaneously by constructing partial off-shell supersymmetry on $S^4$ 
and shown that the localization works well\footnote{
Recently many supersymmetric field theories on some curved backgrounds with rigid SUSY 
have been also constructed in the context of localization method 
for three dimensions \cite
{Kim:2009wb,Kapustin:2009kz,Gang:2009wy,Jafferis:2010un,Hama:2010av,Imamura:2011su,Hama:2011ea,Imamura:2011wg,Alday:2012au,Imamura:2012rq}, 
four dimensions \cite{Nawata:2011un,Ito:2011ea,Nagasaki:2011sh,Hama:2012bg} 
and five dimensions \cite{Hosomichi:2012ek,Kawano:2012sx,Kim:2012av,Terashima:2012ra,Imamura:2012xg}. 
See also \cite{Kallen:2011ny, Ohta:2012ev} for
a twist of rigid supersymmetry and the equivariant localization on Seifert manifolds.
}. 
In this case, 
Killing spinors are modified and the mass terms associated with the curvature 
are introduced in order to close the supersymmetry algebra.
Nevertheless, off-shell maximal SYM in other dimensions on curved background
has not been studied so much
while there have been recently remarkable developments 
for less supersymmetric cases \cite{Festuccia:2011ws,Dumitrescu:2012ha,Liu:2012bi,Samtleben:2012gy}
and from the point of view of holography \cite{Klare:2012gn,Cassani:2012ri}.

If we succeed in such constructions,
we can apply the localization method to the maximal SYM 
and compute exactly some BPS observables in principle. 
Since the (p+1)-dimensional maximal SYM is low-energy effective theory of Dp-branes,
it probably enables us to understand nontrivial problems in superstring theory
via the gauge/gravity duality \cite{Maldacena:1997re,Itzhaki:1998dd}. 
Among them, one of the  interesting class is the  five-dimensional maximal SYM. 
It has been conjectured 
that the six-dimensional $\mathcal{N}=(2,0)$ superconformal field theory compactified on a circle 
is described by the five-dimensional maximal SYM \cite{Lambert:2010iw, Douglas:2010iu}. 
The (2,0)-theory is expected to contain $N^3$-degrees of freedom from the gravity side analysis \cite{Klebanov:1996un}. 
However, intrinsic off-shell description of the (2,0)-theory has not been found yet   
and it is difficult to derive the $N^3$-behavior from the field theory side directly.  
The explicit off-shell Lagrangian description of the SYM on $S^5 $ \cite{Hosomichi:2012ek} 
makes it possible to derive the $N^3$-behavior via the localization \cite{Kallen:2012va, Kallen:2012zn}.   
On the other hand, for  higher dimensional cases,
off-shell supersymmetric actions on most curved spaces 
have not been constructed yet.

In this paper, we construct $d\leq 7$ dimensional maximally supersymmetric Yang-Mills theories 
on a class of curved backgrounds with off-shell supercharges by using the Berkovits method \cite{Berkovits:1993zz}.
In particular, we mainly extend the on-shell supersymmetric action formulated in \cite{Blau:2000xg} 
to be off-shell.
The maximal number of independent off-shell supersymmetries depends on the background
while the number is nine for the flat space.

This paper is organized as follows. 
In section~\ref{sec:flat}, we review the Berkovits construction, 
which is the one of the ten-dimensional super-Yang-Mills theory on the flat space 
with maximally 9 off-shell supercharges.
In section~\ref{sec:curved}, 
we construct the action on curved backgrounds admitting a class of Killing spinors. 
In section~\ref{sec:examples}, we provide some concrete examples.
In section~\ref{another}, we introduce another class of three-dimensional maximal SYM on $S^3$
and the BMN matrix model \cite{Berenstein:2002jq}.
Section~\ref{sec:con} is devoted to a summary and discussions.

\section{Maximal SYM on flat space with off-shell supercharges}
\label{sec:flat}
In this section, we review the Berkovits method \cite{Berkovits:1993zz}, 
which enables us to construct ten-dimensional SYM on the flat space 
with maximally 9 off-shell supercharges.

\subsection{Action and supersymmetric transformation}
Let us start with the on-shell supersymmetric action 
of the ten-dimensional $\mathcal{N}=1$ SYM on ${\R}^{9,1}$ or $\mathbb{R}^{10}$: 
\begin{\eqa}
S_\text{on-shell}
&=& \frac{1}{g_{YM}^2} \int d^{10} x {\rm Tr} \Biggl[ 
    \frac{1}{4}F_{MN} F^{MN}  -\frac{1}{2} \Psi \Gamma^M D_M \Psi
   \Biggr] ,
\end{\eqa}
where the fields take value in the real Lie algebra\footnote{
Here we basically use the notation of ref.~\cite{Pestun:2007rz}. 
See appendix.~\ref{app:clifford} for our conventions on the Clifford algebra.
For the $U(N)$ gauge group, 
we take the generators $T^a$ as anti-hermitian and normalization as ${\rm Tr}(T^a T^b )=-\delta_{ab}$. 
Note that we define the Boltzmann weight as $e^{+S}$.
} and $\Psi$ is the sixteen components Majorana-Weyl spinor.
The action is invariant under the supersymmetric transformation
\begin{\eqa}
\delta_\epsilon^{(\text{on-shell})} A_M  &=& \epsilon \Gamma_M \Psi, \NN \\
\delta_\epsilon^{(\text{on-shell})} \Psi &=& \frac{1}{2} F_{MN}\Gamma^{MN}\epsilon , 
\end{\eqa}
where $\epsilon$ is the ``bosonic`` constant Majorana-Weyl spinor.
It is well-known that the algebra generated by the transformation closes only on-shell
since the off-shell bosonic degrees of freedom differ from the fermionic ones.

Here we add $16-(10-1)=7$ auxiliary fields to the action\footnote{
Here we take $K_m$ to be imaginary.
} \cite{Berkovits:1993zz} as
\begin{\eqa}
S_{10d} &=& \frac{1}{g_{YM}^2} \int d^{10} x {\rm Tr} \Biggl[ 
    \frac{1}{4}F_{MN} F^{MN}   -\frac{1}{2} \Psi \Gamma^M D_M \Psi -\frac{1}{2}K_m K_m
   \Biggr] ,
\end{\eqa}
and modify the transformation as
\begin{\eqa}
\delta_\epsilon^{(10d)} A_M  &=& \epsilon \Gamma_M \Psi, \NN \\
\delta_\epsilon^{(10d)} \Psi &=& \frac{1}{2} F_{MN}\Gamma^{MN}\epsilon  +K_m \nu_m, \NN \\
\delta_\epsilon^{(10d)} K_m  &=& -\nu_m \Gamma^M D_M \Psi ,
\end{\eqa}
where $m=1\cdots 7$.
$\nu_m$ is also bosonic Majorana-Weyl spinor satisfying 
\begin{\eqa}
\epsilon \Gamma^M \nu_m &=& 0, \NN \\
\frac{1}{2} (\epsilon \Gamma_M \epsilon ) \tilde{\Gamma}^M_{\alpha\beta} 
&=& \nu_\alpha^m \nu_\beta^m  +\epsilon_\alpha \epsilon_\beta, \NN \\
\nu_m \Gamma^M \nu_n &=& \delta_{mn} \epsilon \Gamma^M \epsilon . 
\label{spinor_cond}
\end{\eqa}
Note that $\delta_\epsilon$ is fermionic since $\epsilon$ and $\nu_m$ are bosonic.
We can show that the algebra generated by the transformation closes off-shell.

Let us perform dimensional reduction to $d$-dimensions for a later convenience.
Then, we obtain the action of the maximal SYM on $\R^{d-1,1}$ or $\R^{d}$ in the Berkovits formalism as
\begin{\eqa}
S_{{\rm flat}}
&=& \frac{1}{g_{YM}^2} \int d^d x {\rm Tr} \Biggl[ 
    \frac{1}{4}F_{MN} F^{MN}   -\frac{1}{2} \Psi \Gamma^M D_M \Psi -\frac{1}{2}K_m K_m
   \Biggr] .
\label{flat_action}
\end{\eqa}
$A_M$ denote gauge fields and adjoint scalars:
\begin{\eq}
 A_{M=\mu} = A_\mu ,\quad A_{M=A} = \Phi_A ,
\end{\eq}
where $\mu =1\cdots d , A=d+1 \cdots 9,0$.
The field strength is explicitly given by
\begin{\eq}
F_{\mu\nu} = \del_\mu A_\nu -\del_\nu A_\mu +[A_\mu ,A_\nu ] ,\quad
F_{\mu A} = D_\mu \Phi_A ,\quad
F_{AB} = [\Phi_A ,\Phi_B ] ,
\end{\eq}
and the covariant derivative is
\begin{\eq}
D_\mu \Psi = \del_\mu \Psi +[A_\mu ,\Psi] ,\quad
D_A \Psi = [\Phi_A ,\Psi ].
\end{\eq}
This action is invariant under the transformation
\begin{\eqa}
\delta_\epsilon^{({\rm flat})} A_M  &=& \epsilon \Gamma_M \Psi, \NN \\
\delta_\epsilon^{({\rm flat})} \Psi &=& \frac{1}{2} F_{MN}\Gamma^{MN}\epsilon  +K_m \nu_m, \NN \\
\delta_\epsilon^{({\rm flat})} K_m   &=& -\nu_m \Gamma^M D_M \Psi .
\label{flat_d}
\end{\eqa}
If we integrate $K_m$ out, we can realize the well-known on-shell supersymmetric action and transformation.

\subsection{The number of the off-shell supercharges}
In the Berkovits method, 
the supersymmetric transformation (\ref{flat_d}) is generated by
the 8 spinors $\epsilon ,\nu_m$ constrained by the condition (\ref{spinor_cond}).
In order to count the number of the off-shell supercharges in conventional SUSY,
we should rewrite the transformation (\ref{flat_d}) in the language of the single spinor $\epsilon$ 
by solving the constraint (\ref{spinor_cond}).
In \cite{Evans:1994np}, 
the author has constructed the explicit solutions, 
which have 8, 9 and 4 linearly independent spinor parameters $\epsilon$.
In this section we introduce these solutions.

\subsubsection*{Solution with eight off-shell supersymmetries}
Here we impose the restriction to $\epsilon$ as
\begin{\eq}
\Gamma_{09} \epsilon = \epsilon ,
\end{\eq}
giving eight linearly independent supercharges.
Then one can show that the spinors 
\begin{\eq}
\nu_m = \Gamma_{m 8} \epsilon 
\label{eight_sol}
\end{\eq}
solve the constraint (\ref{spinor_cond}).
It is convenient to use the following representation of the Clifford algebra:
\begin{\eqa}
&& \Gamma^0 = \begin{pmatrix} \mathbf{1}_{8\times 8}  & 0 \cr 0 &  \mathbf{1}_{8\times 8}  \end{pmatrix},\quad 
   \Gamma^9 = \begin{pmatrix} \mathbf{1}_{8\times 8}  & 0 \cr 0 & -\mathbf{1}_{8\times 8}  \end{pmatrix}, \NN \\
&& \Gamma^8 = \begin{pmatrix} 0 & \mathbf{1}_{8\times 8}  \cr \mathbf{1}_{8\times 8} & 0  \end{pmatrix},\quad 
   \Gamma^m = \begin{pmatrix} 0 & \lambda_m  \cr -\lambda_m  & 0  \end{pmatrix}, 
\end{\eqa}
where $\lambda_m$ is the anti-symmetric matrix satisfying\footnote{This condition is satisfied by octonions.}
\begin{\eq}
\lambda_m \lambda_n +\lambda_n \lambda_m = -2\delta_{mn} .
\end{\eq}
This representation naturally decomposes ten-dimensional spinors into eigenstates of $\Gamma_{09}$ as
$\chi =(\chi^{(+)}, \chi^{(-)})$ with $\Gamma_{09}\chi^{(\pm )} =\pm \chi^{(\pm )}$.
Then, the solution (\ref{eight_sol}) is rewritten as
\begin{\eq}
\nu_m^{(+)} =\lambda_m \epsilon^{(+)},\quad \nu_m^{(-)} = \epsilon^{(-)} =0.
\end{\eq}

\subsubsection*{Solution with nine off-shell supersymmetries}
There is also a similar solution:
\begin{\eq}
\nu_m^{(+)} = n \lambda_m n \epsilon^{(+)},\quad 
\nu_m^{(-)} = -\lambda_m \epsilon^{(-)} =0,\quad
n \epsilon^{(-)} = \epsilon^{(-)},
\label{nine_sol}
\end{\eq}
where $n_{\alpha \beta} ={\rm diag}(-1,-1,-1,-1,-1,-1,-1,+1)$.
While the projection of $\epsilon^{(-)}$ onto the eigenspinor of $n$ restricts 
$\epsilon^{(-)}$ to one degree of freedom,
the projection does not restrict $\epsilon^{(+)}$. 
Thus, the number of the off-shell SUSY is nine.

\subsubsection*{Solution with four off-shell supersymmetries}
If we take $\epsilon$ and $\nu_m$ as
\begin{\eqa}
&& \Gamma_{45}\epsilon = \Gamma_{67}\epsilon  = \Gamma_{89}\epsilon , \NN \\
&& \nu_1 = \Gamma_{68} \epsilon = -\Gamma_{79}\epsilon ,\quad
   \nu_2 =-\Gamma_{69} \epsilon = -\Gamma_{78}\epsilon ,\NN\\
&& \nu_3 = \Gamma_{84} \epsilon = -\Gamma_{95}\epsilon ,\quad
   \nu_4 =-\Gamma_{85} \epsilon = -\Gamma_{94}\epsilon , \NN \\
&& \nu_5 = \Gamma_{46} \epsilon = -\Gamma_{57}\epsilon ,\quad
   \nu_6 =-\Gamma_{47} \epsilon = -\Gamma_{56}\epsilon , \NN \\
&& \nu_7 =-\Gamma_{45} \epsilon = -\Gamma_{67}\epsilon = -\Gamma_{89}\epsilon ,
\end{\eqa}
then this also solves the constraint (\ref{spinor_cond}).
Since two linearly independent conditions are imposed,
the solution leaves us four off-shell supersymmetries.

\section{Maximal SYM on curved space with off-shell supercharges}
\label{sec:curved}
In this section, 
we construct the $d$-dimensional maximal SYM on a class of curved spaces with off-shell SUSY 
by extending the Berkovits method \cite{Berkovits:1993zz}.

\subsection{Killing spinor}
\label{sec:Killing}
In this paper, 
we construct the maximal SYM, which is invariant under the SUSY generated by Killing spinors satisfying
\begin{\eq}
\nabla_\mu \epsilon = \alpha \tilde{\Gamma}_\mu \Gamma \epsilon ,
\label{Killing_eq}
\end{\eq}
where $\Gamma$ is the product of ``internal`` gamma matrices. 
We denote the number of the internal gamma matrices by $\sharp$.
The on-shell supersymmetric action invariant under the SUSY has been constructed in \cite{Blau:2000xg}.
In order to be consistent with the chirality condition, we take $\sharp$ to be odd below. 
In the Riemannian case, the Killing spinor equation implies the so-called first integrability condition:
\begin{\eq}
R_{\mu\nu} = -4\alpha^2 (-1)^\sharp (\Gamma )^2 (d-1) g_{\mu\nu} .
\label{first_inte}
\end{\eq}
Note that this condition is not sufficient for the existence of the Killing spinors\footnote{
In the pseudo-Riemannian case, the condition (\ref{first_inte}) is neither necessary nor sufficient except for $d=1$.
For details, see \cite{Blau:2000xg}.
}. 

Supersymmetry on curved space admitting a type of Killing spinor including (\ref{Killing_eq}) 
has been studied in ref.~\cite{Klare:2012gn}. 
First, introducing the Dirac operator $D=\Gamma^\mu \nabla_\mu$, 
we can rewrite the Killing spinor equation as
\begin{\eq}
\nabla_\mu \epsilon =\frac{1}{d}\Gamma_\mu D \epsilon .
\end{\eq}
Note that $D\epsilon$ satisfies the following relation from the Killing spinor eq.~(\ref{Killing_eq})
\begin{\eq}
\nabla_\mu (D\epsilon) 
= d \lambda^2 \tilde{\Gamma}_\mu \epsilon 
\quad {\rm with}~~ \lambda^2 =- \alpha^2 (\Gamma )^2 .
\end{\eq} 
By using these relations, one can easily show that
the spinors $\epsilon_\pm =\mp \lambda \epsilon - \frac{1}{d} D \epsilon$ satisfy
\begin{\eq}
\nabla_\mu \epsilon_\pm =\pm \lambda \Gamma_\mu \epsilon_\pm \label{decompose} .
\end{\eq}
The Riemannian manifolds admitting this type of Killing spinor equation have been classified well 
in refs.~\cite{springerlink:10.1007/BF00128299,springerlink:10.1007/BF02102106,springerlink:10.1007/BF00127864}.
For real $\lambda$ case, 
it has been shown to be admitted on non-compact manifolds,
which are a warped product of $\mathbb{R}$ with any $(d-1)$-dimensional manifolds \cite{springerlink:10.1007/BF00128299}.
For purely imaginary $\lambda$ case, the situation becomes somewhat complicated.
The existence of the Killing spinor on the space $\mathcal{M}_d$ implies the existence of 
a covariantly constant spinor on the cone $C(\mathcal{M}_d )$ \cite{springerlink:10.1007/BF02102106}.
For example, the four-dimensional space $\mathcal{M}_4$ should be conformally flat manifold. 
For $d=3$, it is admitted for $\mathcal{M}_3 =S^3$ or its quotients.
In higher dimensions, the Killing spinor is admitted for Sasaki-Einstein manifolds in five dimensions
and nearly K\"{a}hler manifolds in six dimensions, respectively.
In sec.~\ref{sec:examples}, we provide some concrete examples.

\subsection{Construction of action}
Since the Killing spinors on a curved space are 
no longer covariantly constants ($\nabla_\mu \epsilon \neq 0$) generically,
the naive extension of the flat action (\ref{flat_action})
\begin{\eqa}
S_0
&=& \frac{1}{g_{YM}^2} \int d^d x \sqrt{g} {\rm Tr} \Biggl[ 
    \frac{1}{4}F_{MN} F^{MN}   -\frac{1}{2} \Psi \Gamma^M D_M \Psi -\frac{1}{2}K_m K_m
   \Biggr] 
\end{\eqa}
is not invariant under the transformation
\begin{\eqa}
\delta_\epsilon^{(0)} A_M  &=& \epsilon \Gamma_M \Psi, \NN \\
\delta_\epsilon^{(0)} \Psi &=& \frac{1}{2} F_{MN}\Gamma^{MN}\epsilon  +K_m \nu_m, \NN \\
\delta_\epsilon^{(0)} K_m   &=& -\nu_m \Gamma^M D_M \Psi .
\end{\eqa}
Actually the variation of the action under $\delta_\epsilon^{(0)}$ is
\begin{\eqa}
g_{\rm YM}^2 \delta_\epsilon^{(0)} S_0
&=&  \int d^d x \sqrt{g} {\rm Tr} \Biggl[ 
         -\frac{1}{2} F_{NP} (\nabla_\mu  \epsilon )  \Gamma^{NP\mu}  \Psi 
         +F^{\mu N} (\nabla_\mu \epsilon ) \Gamma_N  \Psi  \Biggr] \NN \\
&= & (-1)^{\Gamma +1}  \int d^d x \sqrt{g} {\rm Tr} \Biggl[ 
           \frac{\alpha (d-4)}{2} F_{\mu\nu} (\epsilon \Gamma  \Gamma^{\mu\nu}  \Psi )  
          + \alpha (d-2) F_{\mu A} (\epsilon \Gamma  \Gamma^{\mu A}  \Psi )  \NN \\
&&        ~~~~~~~~~~~~~~~~~~~~~~~~~~~~~~~+ \frac{\alpha d}{2} F_{AB}     (\epsilon \Gamma  \Gamma^{AB}  \Psi )  \Biggr] ,
\end{\eqa}
where $(-1)^\Gamma$ is defined by
\begin{\eq}
\Gamma^T = (-1)^\Gamma \Gamma,
\end{\eq}
in terms of the transposition $\Gamma^T$.
This cannot vanish except for the case $\nabla_\mu \epsilon =0~(\alpha =0)$, namely, Ricci-flat case. 
Therefore in order to obtain desired actions,
we must deform the above action and transformation.
We consider the well-known case for $d=4$ and then the less-known case for $1\leq d\leq 7$.
The properties of SUSY algebra will be investigated in sec.~\ref{sec:algebra}.

\subsubsection{Class 1 ($d=4$)}
Let us consider the following action
\begin{\eqa}
S_1
&=& S_0 +\frac{1}{g_{YM}^2} \int d^d x \sqrt{g} {\rm Tr} \Biggl[ 
   \frac{c_\Phi}{2} \Phi_A \Phi^A   \Biggr] ,
\end{\eqa}
and transformation 
\begin{\eqa}
\delta_\epsilon^{(1)} A_M  &=& \epsilon \Gamma_M \Psi, \NN \\
\delta_\epsilon^{(1)} \Psi &=& \frac{1}{2} F_{MN}\Gamma^{MN}\epsilon  
                              +c \Phi_A \tilde{\Gamma}^A \Gamma \epsilon +K_m \nu_m, \NN \\
\delta_\epsilon^{(1)} K_m  &=& -\nu_m \Gamma^M D_M \Psi  .
\end{\eqa}
The transformation under $\delta_\epsilon^{(1)}$ is
\begin{\eqa}
&& g_{\rm YM}^2 \delta_\epsilon^{(1)} S_1 \NN \\
&=& (-1)^{\Gamma +1}  \int d^3 x \sqrt{g} {\rm Tr} \Biggl[ 
         \frac{\alpha (d-4)}{2} F_{\mu\nu} (\epsilon \Gamma  \Gamma^{\mu\nu}  \Psi )  
         +\left( \alpha (d-2) +c \right) F_{\mu A} (\epsilon \Gamma  \Gamma^{\mu A}  \Psi )  \NN \\
&&       +\left( \frac{\alpha d}{2} +c \right) F_{AB}     (\epsilon \Gamma  \Gamma^{AB}  \Psi )     
      +(-1)^{\Gamma +1} \left( c_\Phi  +c \alpha d (\Gamma )^2  \right) \Phi_A \left( \epsilon \Gamma^A  \Psi \right) 
       \Biggr]. 
\end{\eqa}
The condition for invariance is
\begin{\eq}
\alpha (d-4) =0,\quad  \alpha (d-2) +c =0,\quad  \frac{\alpha d}{2} +c =0,\quad
c_\Phi  +c \alpha d (\Gamma )^2  =0 .
\end{\eq}
This condition has a nontrivial solution only for $d=4$:
\begin{\eq}
c =-2\alpha ,\quad c_\Phi = 8\alpha^2 (\Gamma )^2 .
\end{\eq}
Thus, the action
\begin{\eqa}
S_{4d}
&=& \frac{1}{g_{YM}^2} \int d^4 x \sqrt{g} {\rm Tr} \Biggl[ 
    \frac{1}{4}F_{MN} F^{MN} +4\alpha^2 (\Gamma )^2 \Phi_A \Phi^A \NN \\
&&   ~~~~~~~~~~~~~~~~~~~~~~~~~~-\frac{1}{2} \Psi \Gamma^M D_M \Psi -\frac{1}{2}K_m K_m
   \Biggr] ,
\label{action_4d}
\end{\eqa}
is invariant under the transformation
\begin{\eqa}
\delta_\epsilon^{(4d)} A_M  &=& \epsilon \Gamma_M \Psi, \NN \\
\delta_\epsilon^{(4d)} \Psi &=& \frac{1}{2} F_{MN}\Gamma^{MN}\epsilon  
                               -2\alpha \Phi_A \tilde{\Gamma}^A \Gamma \epsilon +K_m \nu_m, \NN \\
\delta_\epsilon^{(4d)} K_m  &=& -\nu_m \Gamma^M D_M \Psi  .
\label{trans_class1}
\end{\eqa}

\subsubsection*{Remarks}
Since $4\alpha^2 (\Gamma )^2 =\frac{R}{12}$,
we have just recovered the so-called conformal mass term for the scalars.
Actually we can obtain the action (\ref{action_4d}) by performing Weyl transformation 
from the flat action (\ref{flat_action}) for conformally flat spaces \cite{Pestun:2007rz,Nagasaki:2011sh}.
The authors have considered more general (conformal) Killing spinors satisfying
\begin{\eq}
\nabla_\mu \epsilon = \tilde{\Gamma}_\mu \tilde{\epsilon} ,
\label{conformal_Killing}
\end{\eq}
which is nothing but the conformal Killing spinor equation.
In this construction,
the numbers of the on-shell and off-shell supercharges are
thirty-two and eighteen, respectively.

\subsubsection{Class 2 ($d\leq 7 ,\Gamma =\Gamma^{789}$)}
Let us take\footnote{Of course we can arbitrarily take other combinations of three indices from the internal space.} 
$d\leq 7$ and $\Gamma =\Gamma^{789}$.
Let us consider the action
\begin{\eqa}
S_2
&=& S_0 +\frac{1}{g_{YM}^2} \int d^d x \sqrt{g} {\rm Tr} \Biggl[ 
   \frac{c_\Phi}{2} \Phi_A \Phi^A   +\frac{c_\Phi^\prime}{2} \Phi_p \Phi^p 
  +c_Y \epsilon_{pqr} \Phi^p [\Phi^q ,\Phi^r ]  
  +\frac{c_\Psi}{2} \Psi \Gamma^{789} \Psi \Biggr] ,\NN \\
\end{\eqa}
and the transformation 
\begin{\eqa}
\delta_\epsilon^{(2)} A_M  &=& \epsilon \Gamma_M \Psi, \NN \\
\delta_\epsilon^{(2)} \Psi   &=& \frac{1}{2} F_{MN}\Gamma^{MN}\epsilon  +c \Phi_A \tilde{\Gamma}^A \Gamma^{789} \epsilon 
               +c^\prime \Phi_p \tilde{\Gamma}^p \Gamma^{789} \epsilon +K_m \nu_m, \NN \\
\delta_\epsilon^{(2)} K_m     &=& -\nu_m \Gamma^M D_M \Psi  +c_K \nu_m \Gamma^{789} \Psi ,
\end{\eqa}
where $p,q,r =7,8,9$ and $\hat{A}=d+1,d+2,\cdots 6,0$.
The variation under $\delta_\epsilon^{(2)}$ is
\begin{\eqa}
&& g_{YM}^2 \delta_\epsilon^{(2)} S_2 \NN \\
&=& \int d^d x \sqrt{g} {\rm Tr} \Biggl[ 
         \frac{\alpha (d-4) -c_\Psi}{2} F_{\mu\nu} (\epsilon \Gamma^{789} \Gamma^{\mu\nu}  \Psi )  \NN \\
&&  +\left(\alpha  (d-2)  +c +c_\Psi +c^\prime \right) F_{\mu p} (\epsilon \Gamma^{789}  \Gamma^{\mu p}  \Psi )  
    +\left(\alpha  (d-2) +c -c_\Psi  \right) F_{\mu \hat{A}} (\epsilon \Gamma^{789}  \Gamma^{\mu \hat{A}}  \Psi )  \NN \\
&&  +\left( \frac{\alpha d -c_\Psi }{2} +c -3c_Y +c^\prime  \right) F_{pq}  (\epsilon \Gamma^{789} \Gamma^{pq}  \Psi )     
    +\left(\alpha  d+c_\Psi  +2c +c^\prime \right) F_{p\hat{A}}  (\epsilon \Gamma^{789} \Gamma^{p\hat{A}}  \Psi ) \NN \\  
&&  +\left( \frac{\alpha d -c_\Psi }{2} +c \right) F_{\hat{A}\hat{B}}  (\epsilon \Gamma^{789} \Gamma^{\hat{A}\hat{B}} \Psi )     
    \NN \\
&&  +\left( c_\Phi  -c \alpha d   +c_\Psi c -c^\prime \alpha d +c_\Psi c^\prime +c_\Phi^\prime \right)  
                      \Phi_p \left( \epsilon \Gamma^p  \Psi \right) 
    +\left( c_\Phi  -c \alpha d  -c_\Psi c \right)   \Phi_{\hat{A}} \left( \epsilon \Gamma^{\hat{A}}  \Psi \right) \NN \\
&&  +(c_K -c_\Psi ) K_m \Psi \Gamma^{789} \nu_m     \Biggr]  .
\end{\eqa}
Therefore the condition for invariance is
\begin{\eqa}
&& c_\Psi = \alpha(d-4) ,\quad \alpha (d-2) +c +c_\Psi +c^\prime =0,\quad \alpha (d-2) +c -c_\Psi  =0 , \NN \\
&& \alpha  d-c_\Psi  +2c -6c_Y +2c^\prime =0 ,\quad \alpha d +c_\Psi  +2c +c^\prime =0,\quad \alpha d -c_\Psi +2c=0 ,\NN \\
&& c_\Phi  -c \alpha d +c_\Psi c -c^\prime \alpha d +c_\Psi c^\prime +c_\Phi^\prime =0 ,\quad 
  c_\Phi  -c \alpha d -c_\Psi c =0,\quad    c_K =c_\Psi .
\end{\eqa}
This has the following unique nontrivial solution:
\begin{\eqa}
&& c_\Phi = -4\alpha^2 (d-2) ,\quad  c_\Phi^\prime =-4\alpha^2(d-4) ,\quad c_\Psi = \alpha(d-4) ,\quad 
   c_Y = -\frac{2\alpha(d-4)}{3} , \NN \\
&& c =-2\alpha ,\quad c^\prime = -2\alpha(d-4) ,\quad c_K =c_\Psi = \alpha(d-4) .
\end{\eqa}
Thus, the action
\begin{\eqa}
S_2
&=& \frac{1}{g_{YM}^2} \int d^d x \sqrt{g} {\rm Tr} \Biggl[ 
    \frac{1}{4}F_{MN} F^{MN} -2\alpha^2(d-2) \Phi_A \Phi^A  -2\alpha^2(d-4) \Phi_p \Phi^p \NN \\
&&  -\frac{2\alpha(d-4)}{3} \epsilon_{pqr} \Phi^p [\Phi^q ,\Phi^r ]   
    -\frac{1}{2} \Psi \Gamma^M D_M \Psi   +\frac{\alpha(d-4)}{2} \Psi \Gamma^{789} \Psi
   -\frac{1}{2}K_m K_m
   \Biggr] , \NN \\
\end{\eqa}
is invariant under the transformation
\begin{\eqa}
\delta_\epsilon^{(2)} A_M  &=& \epsilon \Gamma_M \Psi, \NN \\
\delta_\epsilon^{(2)} \Psi &=& \frac{1}{2} F_{MN}\Gamma^{MN}\epsilon  -2\alpha \Phi_A \tilde{\Gamma}^A \Gamma^{789} \epsilon 
               -2\alpha(d-4) \Phi_p \tilde{\Gamma}^p \Gamma^{789} \epsilon +K_m \nu_m, \NN \\
\delta_\epsilon^{(2)} K_m  &=& -\nu_m \Gamma^M D_M \Psi  +\alpha(d-4) \nu_m \Gamma^{789} \Psi .
\label{trans_class2}
\end{\eqa}
If we set $d=4 ,\Gamma^{789}\rightarrow \Gamma$, this is same as the one of class 1.

\subsubsection*{Remarks}
Since the action is complex except for $d=4$,
the theory seems to be not reflection positive.
Similar situations have appeared in less supersymmetric gauge theories \cite{Dumitrescu:2012ha}.
Therefore we consider that
our result is their highly supersymmetric version.

For the $d=2$ and purely imaginary $\alpha$ case, 
the scalars $\Phi_7 ,\Phi_8$ and $\Phi_9$ have negative square of mass 
in spite of the positive curvature of the background.
We are suspicious to that the theory for the case is not well-defined.

\subsection{Supersymmetry algebra}
\label{sec:algebra}
In this section, we investigate properties of the SUSY algebras generated 
by the transformations (\ref{trans_class1}) and (\ref{trans_class2}) 
and confirm that these algebras close off-shell.

\subsubsection{Class 1 ($d=4$)}
Let us consider the square of the supersymmetric transformation for each field. 
According to ref.~\cite{Pestun:2007rz}, we obtain
\begin{\eqa}
(\delta_\epsilon^{(4d)})^2 A_\mu  &=&  -v^\nu F_{\nu\mu} -[v^{A} \Phi_{A} ,D_\mu ], \NN\\
(\delta_\epsilon^{(4d)})^2 \Phi_{A}  &=&  -v^\mu D_\mu \Phi_{\hat{A}}-[v^B \Phi_B , \Phi_{A}]-\bar{R}_{AB} \Phi^{B}-\Omega \Phi_A,
                                           \NN \\
(\delta_\epsilon^{(4d)})^2 \Psi  &=&  -v^{\mu}D_\mu \Psi -[v^A \Phi_A , \Psi]
                                      -\frac14( -\bar{R}_{\mu \nu}\Gamma^{\mu\nu}+\bar{R}_{A B}\Gamma^{A B })\Psi-\frac32 \Omega \Psi,
                                        \NN\\
(\delta_\epsilon^{(4d)})^2 K_m  &=&  -v^\mu D_\mu K_m-[v^A \Phi_A , K_m]
                                    -(\nu_{[m}\slashb{D}\nu_{n]})K_n -2 \Omega K_m,
\end{\eqa}
where
\begin{\eqa}
v^M = \epsilon \Gamma^M \epsilon, \quad \bar{R}_{MN}=2 \alpha (\epsilon \tilde{\Gamma}_{MN} \Gamma \epsilon),
      \quad \Omega =2 \alpha (\epsilon \Gamma \epsilon) .
\end{\eqa}
Therefore the square of the supersymmetry transformation can be written as
\begin{\eqa}
(\delta_\epsilon^{(4d)})^2=-L_{v}-G_{\Phi}-\bar{R}-\Omega.
\end{\eqa}
Here $L_{v}$, $G_{\Phi}$ and $\bar{R}$ are a Lie derivative 
in the direction of the (conformal Killing) vector field $v^{\mu}$, 
the gauge transformation generated by the adjoint valued scalar field $\Phi=v^A \Phi_A$ and
a transformation of coordinate or $R$-symmetry transformation, respectively. 
Finally $\Omega$ generates a local dilatation with the parameter $2 \alpha (\epsilon \Gamma \epsilon)$.
Note that the dilatation $\Omega$ vanishes for $\Gamma =\Gamma^{789}$ 
corresponding to the case for $d=4$ of the class 2. 

\subsubsection{Class 2 ($d\leq 7 ,\Gamma =\Gamma^{789}$)}
After some tedious calculations, we can obtain 
\begin{\eqa}
(\delta_\epsilon^{(2)})^2 A_\mu  &=&  -v^\nu F_{\nu\mu} -[v^{A} \Phi_{A} ,D_\mu ], \NN \\
(\delta_\epsilon^{(2)})^2 \Phi_{\hat{A}}  &=&  -v^\nu D_\nu \Phi_{\hat{A}}-[v^B \Phi_B , \Phi_{\hat{A}}]
                                             -\bar{R}_{\hat{A}\hat{B}}  \Phi^{\hat{B}}, \NN \\
(\delta_\epsilon^{(2)})^2 \Phi_p  &=&  -v^\nu D_\nu \Phi_p -[v^B \Phi_B , \Phi_p ]-(d-3)\bar{R}_{p q} \Phi^q, \NN\\
(\delta_\epsilon^{(2)})^2 \Psi  &=&  -v^{\nu}D_\nu \Psi -[v^A \Phi_A , \Psi]
                                     -\frac14( -\bar{R}_{\mu \nu}\Gamma^{\mu\nu}
                                     +\bar{R}_{\hat{A} \hat{B}}\Gamma^{\hat{A}\hat{B}}
                                     +(d-3)\bar{R}_{pq}\Gamma^{pq})\Psi, \NN \\
(\delta_\epsilon^{(2)})^2 K_m  &=&  -v^\mu D_\mu K_m -[v^A \Phi_A , K_m]-(\nu_{[m}\slashb{D}\nu_{n]})K_n,
\end{\eqa}
where
\begin{\eqa}
\bar{R}_{MN}=2 \alpha (\epsilon \tilde{\Gamma}_{MN}\Gamma^{789}\epsilon ) .
\end{\eqa}
Therefore the square of the supersymmetry transformation can be written as
\begin{\eqa}
(\delta_\epsilon^{(2)})^2=-L_{v}-G_{\Phi}-\bar{R}.
\end{\eqa}
Note that the dilatation vanishes automatically in this class as mentioned above. 
Also for the $d=3$ case, R-symmetry transformation of the internal space $p=7,8,9$ vanishes.

\section{Some examples}
\label{sec:examples}
In this section, we provide some concrete examples for the class 2.
As we remarked in sec.~3.2.1,
the class 1 is applicable for confomally flat manifolds.
If we impose the conformal Killing spinor equation (\ref{conformal_Killing}) 
as in  \cite{Pestun:2007rz,Nagasaki:2011sh},
the numbers of the on-shell and off-shell supercharges are
thirty-two and eighteen, respectively.

\subsection{$S^d \ (d\leq 7)$ }
For the class 2, the Killing spinor equation becomes
\begin{\eq}
\nabla_\mu \epsilon = \alpha \tilde{\Gamma}_\mu \Gamma^{789} \epsilon .
\end{\eq}
If we rewrite the equation as (\ref{decompose}) 
and take $\alpha =\frac{i}{2l}$ or $\alpha =-\frac{i}{2l}$, 
\begin{\eq}
\nabla_\mu \epsilon = \frac{i}{2l} \Gamma_\mu  \epsilon \quad {\rm or} \quad
\nabla_\mu \epsilon =-\frac{i}{2l} \Gamma_\mu  \epsilon ,
\label{Killing_sphere}
\end{\eq}
which are the same as the Killing spinor equations on $S^d$ with radius $l$ \cite{Lu:1998nu}.
Actually, from the first integrability condition (\ref{first_inte}), the curvature is given by
\begin{\eq}
R = \frac{d(d-1)}{l^2} ,
\end{\eq}
which is nothing but the curvature of $S^d$.
Since the equation (\ref{Killing_sphere}) does not reduce the degrees of freedom of $\epsilon$,
the number of ``on-shell'' SUSY is sixteen.
If we consider the solution (\ref{nine_sol}) in the same way for the flat space,
then one can easily show that the number of ``off-shell'' SUSY is nine.

\subsection{$AdS_d \ (d\leq 7)$ }
If we set $\alpha =\frac{1}{2l}$ or $\alpha =-\frac{1}{2l}$, 
the Killing spinor equation becomes the same as the one on $AdS_d$ with radius $l$ \cite{Lu:1998nu}.
From the first integrability condition (\ref{first_inte}), the curvature is given by
\begin{\eq}
R = -\frac{d(d-1)}{l^2} ,
\end{\eq}
which is the curvature of $AdS_d$.
Similarly for the $S^d$ case,
the numbers of the on-shell and off-shell supercharges are sixteen and nine, respectively.

\subsection{$S^3 /\mathbb{Z}_n $ }
The lens space $S^3 /\mathbb{Z}_n $ is obtained by $\mathbb{Z}_n$ identification of $S^3$,
which is contained in the subgroup of 
the isometry group $U(1)_R \subset  SU(2)_R\subset SU(2)_L \times SU(2)_R$ \cite{Alday:2012au}.
Since the Killing spinors on $S^3$ are constants along $SU(2)_L$ or $SU(2)_R$,
the latter is also the Killing spinor on $S^3 /\mathbb{Z}_n $.
Thus, the numbers of the on-shell and off-shell supercharges are eight and at least four, respectively.

\section{Another class of 3d $\mathcal{N}=8$ SYM on $S^3$ and the BMN matrix model}
\label{another}
In this section, we construct another class of maximal SYM on $S^3$ 
and  the BMN matrix model \cite{Berenstein:2002jq} with off-shell SUSY.
These theories are obtained by dimensional reducing 4d $\mathcal{N}=4$ SYM 
on $\R\times S^3$.

\subsection{Another class of 3d $\mathcal{N}=8$ SYM on $S^3$}
Let us start with 4d $\mathcal{N}=4$ SYM on $\R\times S^3$.
It is well known that $\R\times S^3$ is the conformally flat manifold 
since
\begin{\eqa}
ds_{\R^4}^2
&=& dr^2 +r^2 d\Omega_3^2 \NN \\
&=& e^{\frac{2}{l}\tau } ( d\tau^2 +l^2 d\Omega_3^2 ) \quad ( r=l e^{\frac{\tau}{l}} ) \NN \\ 
&=& e^{\frac{2}{l}\tau} ds_{\R\times S^3}^2 .
\end{\eqa}
This means that the theory on $\R\times S^3$ is obtained by the Weyl transformation 
$g_{\mu\nu}^{\R\times S^3} = e^{2\omega} g_{\mu\nu}^{\R^4}$ with $\omega =-\frac{\tau}{l}$ from the theory on $\R^4$.
Since the curvature of $\R\times S^3$ is $\frac{6}{l^2}$,
we obtain
\begin{\eqa}
S_{\R\times S^3}
&=& \frac{1}{g_{YM}^2} \int d\tau d\Omega_3 {\rm Tr} \Biggl[ 
    \frac{1}{4}F_{\mu\nu}F^{\mu\nu} +\frac{1}{2}D_\mu \Phi_A D^\mu \Phi^A
    +\frac{1}{2l^2} \Phi_A \Phi^A  \NN\\
&&  +\frac{1}{4} [\Phi_A ,\Phi_B ] [\Phi^A ,\Phi^B ] 
    -\frac{1}{2}\Psi \Gamma^\mu D_\mu \Psi
    -\frac{1}{2}\Psi \Gamma^A [ \Phi_A , \Psi ] -\frac{1}{2}K_m K_m
   \Biggr] , \label{another-action}
\end{\eqa}
which is invariant under the transformation
\begin{\eqa}
\delta_\epsilon^{(\R\times S^3 )} A_M  &=& \epsilon \Gamma_M \Psi, \NN \\
\delta_\epsilon^{(\R\times S^3 )} \Psi &=& \frac{1}{2} F_{MN}\Gamma^{MN}\epsilon  
                                          +\frac{1}{2}\Gamma^{\mu A} \Phi_A \nabla_\mu \epsilon +K_m \nu_m, \NN \\
\delta_\epsilon^{(\R\times S^3 )} K_m  &=& -\nu_m \Gamma^M D_M \Psi  . \label{another-tr}
\end{\eqa}
The conformal Killing spinor satisfies
\begin{\eqa}
\nabla_\mu \epsilon 
= \tilde{\Gamma}_\mu \tilde{\epsilon} ,\quad
\nabla_\mu \tilde{\epsilon} 
= -\frac{1}{8l^2} \Gamma_\mu \epsilon ,
\end{\eqa}
which are solved as
\begin{\eqa}
\epsilon = e^{\frac{1}{2}\omega} (\epsilon_s +x^\mu \tilde{\Gamma}_\mu \epsilon_c ) ,\quad
\tilde{\epsilon}
&=& e^{\frac{1}{2}\omega} \left( 
                \frac{1}{2} e^{-\omega} \epsilon_c           
                       -\frac{1}{2l^2} e^{\omega}  \Gamma^a  x_a   \epsilon_s  \right)  ,
\end{\eqa}
where $\epsilon_s$ and $\epsilon_c$ are constant spinors.
The supersymmetry algebra generated by the spinors is \cite{Pestun:2007rz}
\begin{\eqa}
(\delta_\epsilon^{(\R\times S^3 )})^2 A_\mu  &=&  -v^\nu F_{\nu\mu} -[v^{A} \Phi_{A} ,D_\mu ], \NN\\
(\delta_\epsilon^{(\R\times S^3 )})^2 \Phi_{A}  &=&  -v^\mu D_\mu \Phi_{\hat{A}}-[v^B \Phi_B , \Phi_{A}]-\bar{R}_{AB} \Phi^{B}-\Omega \Phi_A,
                                           \NN \\
(\delta_\epsilon^{(\R\times S^3 )})^2 \Psi  &=&  -v^{\mu}D_\mu \Psi -[v^A \Phi_A , \Psi]
                                      -\frac14( -\bar{R}_{\mu \nu}\Gamma^{\mu\nu}+\bar{R}_{A B}\Gamma^{A B })\Psi-\frac32 \Omega \Psi,
                                        \NN\\
(\delta_\epsilon^{(\R\times S^3 )})^2 K_m  &=&  -v^\mu D_\mu K_m-[v^A \Phi_A , K_m]
                                    -(\nu_{[m}\slashb{D}\nu_{n]})K_n -2 \Omega K_m ,
\end{\eqa}
where
\begin{\eqa}
v^M = \epsilon \Gamma^M \epsilon, \quad \bar{R}_{MN}=2 (\epsilon \tilde{\Gamma}_{MN} \tilde{\epsilon}),
      \quad \Omega =2 (\epsilon \tilde{\epsilon}) .
\end{\eqa}
Since the four-dimensional $\mathcal{N}=4$ SYM is the conformal field theory,
the dilatation $\Omega$ is the one of the symmetry and therefore harmless.
However, the nonzero dilatation becomes harmful after the dimensional reduction. 
Thus, we should impose the constraint $\epsilon\tilde{\epsilon}=0$, 
which reduces the degrees of freedom of $\epsilon$ to sixteen.

In order to derive the action of 3d $\mathcal{N}=8$ SYM on $S^3$, 
we apply dimensional reduction along $\R$-direction.
Expanding the gauge field as $A=\phi d\tau +A^i e^i$, we derive
the field strength as 
\begin{eqnarray}
F 
&=& dA +A\wedge A \NN\\
&=& \left( -i\frac{2}{l} (J_i \phi ) +  (\partial_\tau A_i ) +[\phi ,A_i ] \right) d\tau \wedge e^i \NN \\ 
&&  +\frac{1}{2}\epsilon_{ijk} \left\{ \frac{2}{l} i\epsilon_{klm}J_l A_m +\frac{2}{l}A_k 
                                      +\frac{1}{2}\epsilon_{klm} [ A_l ,A_m ] \right\}    e^i \wedge e^j , \label{reducing-gauge}
\end{eqnarray}
where $e^i \ (i=1,2,3)$ is the left-invariant 1-form on $S^3$ and $J_i$ is the Killing vector.
The covariant derivative of the fermion is rewritten as 
\begin{eqnarray}
\Gamma^\mu D_\mu \Psi
&=& \Gamma^\tau D_\tau \Psi +\frac{2i}{l} \Gamma^i J_i \Psi +\frac{3i}{4l}\Gamma^{123}\Psi +\Gamma^i [ A_i ,\Psi ] .
\end{eqnarray}
Applying dimensional reduction along the $\tau$-direction, we obtain 
the following action different from the class 2:
\begin{\eqa}
S_3
&=& \frac{1}{g_{YM}^2} \int d\Omega_3 {\rm Tr} \Biggl[ 
   \frac{1}{2} (\tilde{D}_i \phi )^2 
   +\frac{1}{2} f_{ij} f^{ij} +\frac{1}{2}\tilde{D}_i \Phi_A \tilde{D}^i \Phi^A
    +\frac{1}{2l^2} \Phi_A \Phi^A +\frac{1}{2}[\phi ,\Phi_A ][\phi ,\Phi^A ] \NN\\
&&   +\frac{1}{4} [\Phi_A ,\Phi_B ] [\Phi^A ,\Phi^B ] 
    -\frac{1}{2}\Psi \Gamma^i \tilde{D}_i \Psi -\frac{3i}{8l} \Psi \Gamma^{123}\Psi
    -\frac{1}{2}\Psi \Gamma^\tau [ \phi , \Psi ]  
   -\frac{1}{2}\Psi \Gamma^A [ \Phi_A , \Psi ] -\frac{1}{2}K_m K_m
   \Biggr] , \NN \\ 
\end{\eqa}
where 
\begin{\eqa}
\tilde{D}_i \phi = i\frac{2}{l} J_i \phi +[A_i ,\phi ] ,\quad
f_{ij} = \frac{1}{2} \epsilon_{ijk} 
        \left( \frac{2}{l} i\epsilon_{klm}J_l A_m +\frac{2}{l}A_k  +\frac{1}{2}\epsilon_{klm} [\ A_l ,A_m\ ] \right) .
\end{\eqa}
This action is invariant under the transformation
\begin{\eqa}
\delta_\epsilon^{(3)} A_i   &=& \epsilon \Gamma_i \Psi, \quad
\delta_\epsilon^{(3)} \phi   = \epsilon \Gamma_\tau \Psi , \quad
\delta_\epsilon^{(3)} \Phi_A = \epsilon \Gamma_A   \Psi , \NN \\
\delta_\epsilon^{(3)} \Psi   &=&  \tilde{D}_i \phi \Gamma^{i\tau}\epsilon +\frac{1}{2} f_{ij} \Gamma^{ij}\epsilon 
              +\tilde{D}_i \Phi_A \Gamma^{i A}\epsilon +[\phi ,\Phi_A ] \Gamma^{\tau A}\epsilon
             +\frac{1}{2}[\Phi_A ,\Phi_B ]\Gamma^{AB} \epsilon -2\Phi_A \tilde{\Gamma}^A \tilde{\epsilon} +K_m \nu_m ,\NN \\
\delta_\epsilon^{(3)}K_m     &=& -\nu_m \Gamma^i \tilde{D}_i \Psi -\frac{3i}{4l}\nu_m \Gamma^{123}\Psi 
             -\nu_m \Gamma^\tau [\phi , \Psi ]   -\nu_m \Gamma^A [\Phi_A , \Psi ] . \label{reducing-tr}
\end{\eqa}
One of the main differences from the class 2 is the R-symmetry:
$SO(6)_R$ in this class, while $SO(4)_R \times SO(3)_R$ in the class 2.
Also, note that this action is real and therefore reflection positive different from the class 2.
The numbers of the on-shell and off-shell supercharges are sixteen and nine, respectively.

Similarly, we can generate three different maximal SYM's on $S^2$.
One is directly generated by the class 2 construction.
The others are constructed by the dimensional reduction along the $S^1$-fibre direction of the two different SYM's on $S^3$. 

\subsection{The BMN matrix model}
The action of the BMN matrix model \cite{Berenstein:2002jq} is obtained 
by dimensional reduction of the $\mathcal{N}=4$ SYM on $\mathbb{R}\times S^3$
along the $S^3$-direction \cite{Kim:2003rza}.
In the same way as (\ref{reducing-gauge})-(\ref{reducing-tr}) expanding the gauge field as $A=A_\tau d\tau+X^i e^i$, we obtain 
\begin{\eqa}
S_{\text{BMN}} 
&=&  \frac{1}{g_{\text{YM}}^2} \int d\tau \Tr \Big{[}
      \frac12 D_\tau X_i D^\tau X^i +\frac12 \Big{(} \frac{2}{l}X_i +\frac12 \epsilon_{ijk}[X_j,X_k] \Big{)}^2 \NN \\
&&   +\frac12 D_\tau \Phi_A D^\tau \Phi^A +\frac{1}{2l^2}\Phi_A \Phi^A 
     +\frac12 [X_i , \Phi_A][X^i, \Phi^A] +\frac14 [\Phi_A, \Phi_B][\Phi^A, \Phi^B] \NN \\
&&   -\frac12 \Psi \Gamma^\tau D_\tau \Psi 
     -\frac{3i}{8l}\Psi \Gamma^{123}\Psi -\frac12 \Psi \Gamma^i [X_i , \Psi] 
     -\frac12 \Psi \Gamma^A[\Phi_A, \Psi] -\frac12 K_m K_m \Big{]} .
\end{\eqa}
This action is invariant under the transformation
\begin{\eqa}
\delta_\epsilon^{(\text{BMN})} A_\tau   &=& \epsilon \Gamma_\tau \Psi, \quad
\delta_\epsilon^{(\text{BMN})} X_i   = \epsilon \Gamma_i \Psi , \quad 
\delta_\epsilon^{(\text{BMN})} \Phi_A = \epsilon \Gamma_A   \Psi , \NN \\
\delta_\epsilon^{(\text{BMN})} \Psi   &=&  D_\tau X_i \Gamma^{\tau i}\epsilon +\frac{1}{2} f_{ij} \Gamma^{ij}\epsilon 
              +D_\tau \Phi_A \Gamma^{\tau A}\epsilon \NN \\ 
&&            +[X_i ,\Phi_A ] \Gamma^{i A}\epsilon+\frac{1}{2}[\Phi_A ,\Phi_B ]\Gamma^{AB} \epsilon 
             -2\Phi_A \tilde{\Gamma}^A \tilde{\epsilon} +K_m \nu_m , \NN \\
\delta_\epsilon^{(\text{BMN})}K_m     &=& -\nu_m \Gamma^\tau D_\tau \Psi -\frac{3i}{4l}\nu_m \Gamma^{123}\Psi 
             -\nu_m \Gamma^i \tau [X_i , \Psi ]   -\nu_m \Gamma^A [\Phi_A , \Psi ] . 
\end{\eqa}
 
\section{Conclusion}
\label{sec:con}
We have constructed the $d\leq 7$ dimensional maximally supersymmetric Yang-Mills theories 
on a class of curved backgrounds with the off-shell supercharges.
The class 1 admits the nontrivial rigid supersymmetry 
if and only if the space dimension is four. 
In this case, the theory is conformally equivalent to the one on the flat space and
we can actually recover the action by performing Weyl transformation from the flat action
as in  \cite{Pestun:2007rz,Nagasaki:2011sh}.
The class 2 is more interesting. 
This class is admitted for the space dimension $d\leq 7$ and contains the various concrete examples:
we have explicitly shown that $S^{d}$ and
$AdS_{d}$ are involved in the class and the number of the off-shell supersymmetries is nine. 
By orbfolding $S^3$, the lens space $S^3/\Z_n$ preserves at least four off-shell supercharges. 
This class also involves Sasaki-Einstein and nearly K\"{a}hler manifolds 
for $d=5$ and $d=6$, respectively \cite{springerlink:10.1007/BF00128299,springerlink:10.1007/BF02102106,springerlink:10.1007/BF00127864}
although we did not concretely investigate these cases.
 
In our off-shell construction, supersymmetries are realized linearly. 
Thus it is transparent what types of  the bosonic symmetries are generated 
by the commutator of the supersymmetric transformations.
Especially we emphasize that the dilatation symmetry is not generated in the class 2: 
this is a necessary condition for existence of supersymmetries of non-conformal field theories at quantum level.  
In general, higher-dimensional field theories are not well-defined at quantum level. 
However when off-shell supercharges $Q$ persist at quantum level, 
the localization formula  make it possible to give a constructive formulation of such quantum theories 
at least in $Q$-closed sector. 
It is interesting to study quantum effects and work out localization procedures in these classes
to challenge unsolved problems in superstring/M-theory \cite{Itzhaki:1998dd}.
 
Extension to other classes of backgrounds would be also interesting.
For instance, we could twist by a line bundle $L$ 
and consider the following Killing spinor equation as \cite{Dumitrescu:2012ha} 
\begin{\eq}
(\nabla_\mu -i\tilde{A}_\mu ) \epsilon = -iV_\mu \epsilon -iV^\nu \Gamma_{\mu\nu} \epsilon ,
\end{\eq}
where $\tilde{A}_\mu$ is a connection on $L$ and $V^\mu$ is a smooth, conserved vector field.
Furthermore, extension to backgrounds with torsion \cite{Ito:2012hs} and pseudo-Riemannian manifolds \cite{deMedeiros:2012sb}
would be also illuminating.

\subsection*{Acknowledgment}
The authors would like to thank M.~Hanada, Y.~Imamura, S.~Matsuura, S.~Mizoguchi, F.~Sugino and S.~Yamaguchi
for valuable discussions.
M.H. is supported by Grant-in-Aid for JSPS fellows (No.22-2764).

\appendix
\section{Clifford algebra}
\label{app:clifford}
Let $\gamma^M$ be the standard gamma matrix in the 10d Minkowski space whose metric is given by
$ds^2 = -dx_0^2 +dx_1^2 +\cdots +dx_9^2$.
Then, $\gamma^M$ satisfies the standard anti-commutation relation:
\begin{\eq}
\{ \gamma^M ,\gamma^N \} =2g^{MN} ,
\end{\eq}
where $M,N=1,\cdots,9,0$ and $\gamma^M$ is $32 \times 32$ matrix.
Since we now consider even dimension, we have the chiral operator
\begin{\eq}
\gamma^{11} = \gamma^1 \gamma^2 \cdots \gamma^9 \gamma^0
\end{\eq}
and $\gamma^M$ reverse chirality.
Then, we can write $\gamma^M$ as
\begin{\eq}
\gamma^M 
= \begin{pmatrix} 
0         & \tilde{\Gamma}^M \cr 
\Gamma^M  &    0             \cr
\end{pmatrix}.
\end{\eq}
Then, the ``half'' gamma matrices satisfies the following anti-commutation relation:
\begin{\eq}
 \tilde{\Gamma}^{\{ M}  \Gamma^{N \}}        = g^{MN},\quad
        \Gamma^{\{ M }  \tilde{\Gamma}^{N\}} = g^{MN}.
\end{\eq}
And we choose $\Gamma^M$ and $\tilde{\Gamma}^M$ to be symmetric.
\\

Here we mention some useful formula.
\begin{itemize}
\item Triality identity:
\begin{\eqa}
(\Gamma_M)_{\alpha_1 \{ \alpha_2}(\Gamma^M)_{\alpha_3 \alpha_4 \}}=0
\end{\eqa}
where $\alpha_1,\alpha_2,\alpha_3,\alpha_4=1,\cdots,16$.

\item Fierz identity for $d=10$ Majorana-Weyl spinors with same chirality:
\begin{\eqa}
(\epsilon_1 A \epsilon_2)(\epsilon_3 B \epsilon_4)
&=&  \frac{1}{16}\Big{[}
      (\epsilon_1  \epsilon_4 )( \epsilon_3 B A \epsilon_2)
      -\frac12 (\epsilon_1 \tilde{\Gamma}_{MN} \epsilon_4)(\epsilon_3 B \tilde{\Gamma}^{MN} A \epsilon_2 ) \NN \\
&&~~~~~~    +\frac{1}{4!}( \epsilon_1 \tilde{\Gamma}_{MNKL} \epsilon_4 )( \epsilon_3 B \tilde{\Gamma}^{MNKL} A \epsilon_2)
     \Big{]}
\end{\eqa}
where $\epsilon_1, \epsilon_2, \epsilon_3, \epsilon_4$ are real spinors with 16 components, and A and B are $16 \times 16$ matrices.

\item Other useful formula:
\begin{\eqa}
&& \tilde{\Gamma}^{NP} \Gamma^M = \Gamma^{NPM} -2g^{M[N} \Gamma^{P]} , \quad 
   \tilde{\Gamma}_\rho  \Gamma^{NP\rho}
  =\tilde{\Gamma}_\rho ( \tilde{\Gamma}^{NP} \Gamma^\rho +2g^{\rho [N} \Gamma^{P]} ) ,\NN \\
&& \tilde{\Gamma}_\rho \tilde{\Gamma}^{\mu\nu} \Gamma^\rho = (d-4)\Gamma^{\mu\nu} ,\quad 
   \tilde{\Gamma}_\rho \tilde{\Gamma}^{\mu A} \Gamma^\rho =  (d-2)\Gamma^{\mu A}   ,\quad
   \tilde{\Gamma}_\rho \tilde{\Gamma}^{AB} \Gamma^\rho = d\Gamma^{AB} .
\end{\eqa}
\end{itemize}

\bibliographystyle{JHEP}

\bibliography{localization,curved_SUSY}
\end{document}